\documentclass[]{article}
\usepackage{float}
\usepackage[caption = false]{subfig}
\usepackage[final]{graphicx}
\usepackage[a4paper, total={6in, 8in}]{geometry}
\usepackage{booktabs}
\usepackage{longtable}
\usepackage{natbib}
\usepackage{xcolor}
\usepackage{url}
\usepackage[figuresright]{rotating}

\begin{document}
	
\begin{titlepage}
   \begin{center}
       \vspace*{1cm}
\large
       \textbf{Dynamic Updating of Clinical Survival Prediction Models in a Rapidly Changing Environment}
       \vspace{0.5cm}

by        
            
%       \vspace{2.5cm}
       
\normalsize
       \textbf{Kamaryn Tanner$^{1}$,
       		Ruth H Keogh$^{1}$,
       		Carol AC Coupland$^{2,3}$,\\
       		Julia Hippisley-Cox$^{2}$,
       		Karla Diaz-Ordaz$^{4}$ \\}

      % \vfill

       \vspace{0.8cm}

\small
            
	$^{1}$Dept of Medical Statistics, London School of Hygiene and Tropical Medicine, London WC1E 7HT, UK \\

	$^{2}$Nuffield Department of Primary Health Care Sciences, University of Oxford, Oxford OX2 6HT, UK\\

	$^{3}$Centre for Academic Primary Care, School of Medicine, University of Nottingham, Nottingham NG7 2UH, UK\\

	$^{4}$Dept of Statistical Science, University College London, London WC1E 6BT, UK

	\vspace{1.2cm}
	
	12 April 2023
            
   \end{center}
\end{titlepage}

\begin{abstract}
Over time, the performance of clinical prediction models may deteriorate due to changes in clinical management, data quality, disease risk and/or patient mix. Such prediction models must be updated in order to remain useful.  Here, we investigate methods for discrete and dynamic model updating of clinical survival prediction models based on refitting, recalibration and Bayesian updating. In contrast to discrete or one-time updating, dynamic updating refers to a process in which a prediction model is repeatedly updated with new data. Motivated by infectious disease settings, our focus was on model performance in rapidly changing environments. We first compared the methods using a simulation study. We simulated scenarios with changing survival rates, the introduction of a new treatment and predictors of survival that are rare in the population. Next, the updating strategies were applied to patient data from the QResearch database, an electronic health records database from general practices in the UK, to study the updating of a model for predicting 70-day covid-19 related mortality. We found that a dynamic updating process outperformed one-time discrete updating in the simulations. Bayesian dynamic updating has the advantages of making use of knowledge from previous updates and requiring less data compared to refitting.

\end{abstract}

\clearpage
\section{Introduction}

Clinical prediction models are widely used in medicine to provide patients and clinicians with information about the predicted risk of an outcome, to guide treatment plans, and to identify high-risk groups. Although the best of these models go through a thorough development and validation process, performance may deteriorate over time as clinical practices change, mortality or disease risk in the population changes and/or the patient mix shifts \citep{Steyerberg2009, Moons2012, Hickey2013}. After evidence of poor performance, a common response is to repeat the model development by fitting a new model to a new set of data \citep{Janssen2008}. However, the resulting `new' model fails to incorporate knowledge learned in the initial development process, may not include the same predictors and could be confusing for end-users of the original model \citep{Moons2012}. A technique for updating the existing prediction model rather than redeveloping a new one can alleviate these issues.

A variety of methods have been proposed and studied for updating clinical prediction models with the majority having been applied to binary outcome models based on logistic regression. Previous studies have compared recalibration, refitting (with or without shrinkage), Bayesian methods and testing procedures for selecting the `best’ updating method across a variety of scenarios \citep{Janssen2008, Hickey2013, Sim2016, Siregar2016, Vergouwe2016, Su2018, Davis2019, Davis2020, Schnellinger2021, Riley2021, Feng2022}. No single updating method was best across these studies. Rather, the best updating method was found to depend on sample size of the updating data, event rate, model complexity and whether associations of the predictors with the outcome had changed \citep{Janssen2008, Siregar2016, Vergouwe2016, Davis2019, Riley2021}. In particular, refitting was found to overfit the new data and yield unstable coefficient estimates in parametric models when the updating sample size was small or the number of events was low \citep{Hickey2013, Siregar2016, Vergouwe2016, Schnellinger2021}. Recalibration performed as well as or better than refitting, particularly when predictor relationships were not changing over time \citep{Janssen2008, Su2018, Booth2020, Schnellinger2021}. Bayesian updating methods showed good predictive performance and may produce smoother updates to model coefficients than refitting \citep{Hickey2013, Siregar2016, Su2018}.

Once a prediction model has been updated, it becomes susceptible to performance deterioration again due to evolution of treatments or the disease itself, changes to the affected population, repeated exposures, or quality of the data available for its implementation. Rather than performing updates as a one-time or discrete update, a strategy for continued updating can combat this. A dynamic updating strategy refers to an approach for updating a model at multiple times in the future when new data becomes available \citep{Jenkins2021, Schnellinger2021}. The time period between updates may be fixed or variable and may be as short as the time it takes to receive one new data point or based on a fixed passage of time, e.g. 1 month, 1 quarter or 1 year. Although dynamic updating strategies offer many benefits, their use is still limited because of difficulties in implementing and resourcing such a strategy, obtaining the necessary data and communicating the frequent changes to a prediction model \citep{Hickey2013, Jenkins2021}. Further, the study of updating strategies for time-to-event outcomes has been limited. Clinical examples of updating survival models include the recalibration of Cox proportional hazards models used to predict 30-day survival and 6-month independence after acute stroke by  \citet{Sim2016} and recalibration of Cox models used for predicting survival after a particular treatment for hepatocellular carcinoma by \citet{Cucchetti2021}. 

In this study, our primary aim is to assess methods for dynamic model updating of clinical survival prediction models. We were motivated by the covid-19 pandemic where mortality rates changed over time and new vaccines were introduced to the population. Section 2 provides background on the illustrative example, the QCOVID series of survival prediction models  \citep{Clift2020, Hippisley-Cox2021}. We investigate both one-time and dynamic updating using recalibration, refitting and Bayesian dynamic updating of time-to-event models. We consider data sources where updated data is additional follow-up time on the same individuals and also where updated data refers to data from a new cohort of individuals. These methods are described in section 3. We use a simulation study, presented in section 4, to investigate the performance of multiple updating methods with a focus on calibration, discrimination and variability of hazard ratio estimates to evaluate the performance. In section 5, we illustrate these methods using the QResearch database of electronic health care patient data \citep{QResearch2022} to predict the risk of contracting and dying from covid-19.  We finish with a discussion in Section 6.

\section{Motivating example: a covid-19 clinical prediction model}

Since the beginning of the covid-19 pandemic in early 2020, numerous clinical prediction models have been proposed to assist both clinical decision making and policymakers \citep{Wynants2020}. Among them, the QCOVID series of risk prediction algorithms were developed to help identify those most at risk of getting infected and then dying due to covid-19 based on individual characteristics such as age, sex and long-standing illnesses \citep{Clift2020, Hippisley-Cox2021}. The original QCOVID model was used to prioritise people for vaccination and to inform the government’s shielding list \citep{Clift2020}. It was later refit with data from the second pandemic wave in the UK (QCOVID2) and extended to account for covid-19 vaccination (QCOVID3) \citep{Hippisley-Cox2021}. These models were developed in a rapidly changing environment in terms of availability of vaccines, infection prevalence, levels of immunity in the population, new variants and availability of treatments.

Motivated by the need to keep prediction models such as QCOVID up to date to provide accurate predictions of risk, we apply different updating strategies to a model for prediction of the risk of catching and dying from covid-19 (details found in Section \ref{sec:exMeths}). We use a subset of QResearch data, the same database used to develop the QCOVID models. QResearch is an anonymised database of health records from GP practices throughout the UK that contains historical and current information on over 35 million individuals \citep{QResearch2022}.

\section{Methods for model updating and assessment}

\subsection{Dynamic model updating}
Dynamic model updating is a process whereby a prediction model is repeatedly updated with new information. The process begins during period $u$=0 with an original model $M_0$ fit using a development dataset $D_0$. Dataset $D_0$ contains information from time $t_{-1}$ through time $t_0$. After time $t_0$, new information becomes available during period $u$=1. This new data can be collected to form dataset $D_1$ which contains information from ($t_0$, $t_1$]. The length of this interval may be hours, days, weeks, or months or it could be so small that it only includes the next data point. Using this new dataset, $D_1$, out-of-sample survival predictions are made at a clinically relevant prediction horizon, $v$, using model $M_0$, and performance is measured (see Section \ref{sec:ModAssess}). Model $M_0$ is then updated with the information in $D_1$ and the process repeats.  New data is collected $D_u$, predictions are made using the model $M_{u-1}$, performance is measured and, finally, model $M_{u-1}$ is updated with data $D_u$ resulting in model $M_u$.

\subsection{Performance assessment}
\label{sec:ModAssess}

Predictive performance of the original and updated models can be assessed on out-of-sample predictions by discrimination, calibration and overall performance \citep{Steyerberg2009, McLernon2023}. In time-to-event modelling, given a pair of individuals with known survival times, discrimination refers to the model's ability to assign a greater survival probability to the one who survived longer. We use an inverse probability of censoring weighted (IPCW) C-index to account for censoring \citep{Gerds2013}. Calibration assesses how well predicted outcomes match observed outcomes and we measure weak calibration using calibration intercept and slope \citep{Royston2013, Crowson2016, VanCalster2019}. We use an IPCW Brier score, which is the mean-squared error for binary outcomes and predictions that are probabilities, to measure overall performance \citep{Graf1999}.

\subsection{Methods for model updating}
\label{sec:MethUpdate}

In this section, we review three clinical prediction model updating methods: intercept recalibration, refitting and Bayesian updating and discuss their use in the context of updating a survival prediction model. We will use the term ``no updating" to refer to retaining the original model ($M_0$) without using new data to update it. We assume the original model to be updated is a Cox proportional hazards model \citep{Cox1972}, which can be written:
\begin{equation}\label{eqn:Cox}
	h_i(t \mid X_i) = h_0(t) \exp(\beta^TX_i)
\end{equation}
where $h_i(t \mid X_i)$ is the hazard for the individual $i$ at time $t$, $h_0(t)$ is the baseline hazard, $X_i$ is a vector of time-fixed covariates and $\beta$ is a vector of parameters to be estimated (the log hazard ratios). An estimate of the predicted survival probability for person $i$ can be computed using:
\begin{equation}
	S_i(t \mid X_i) = \exp\lbrace -\hat{H}_0(t) \exp ( \hat{\beta}^TX_i ) \rbrace 
\end{equation}    
where $\hat{H}_0(t)$ is Breslow's estimate of the cumulative baseline hazard \citep{Breslow1972}. Note that with each update, we reset time $t$ to $t=0$ to compute the survival probabilities.

\subsubsection{Intercept Recalibration}
For a Cox proportional hazards model, recalibration of the intercept refers to re-estimating the baseline hazard using the new data while holding the log hazard ratios estimated on the original dataset, $\hat{\beta}$, constant. In period $u$, we first calculate the linear predictor, $\eta_{i,u}$, for the new data using $\hat{\beta}$ from the original model and covariates $X_{i,u}$ from the new data. A Cox model is fit with $\eta_{i,u} = \hat{\beta}^TX_{i,u}$ as the only covariate and its coefficient $\beta_{\eta}$ is fixed at 1. We then estimate the cumulative baseline hazard $H_{0,u}(t)$ at the prediction horizon by setting the value of the linear predictor to zero. Recalibrated survival predictions are computed on the new data as:
\begin{equation}
	S_{i,u}(t \mid X_{i,u}) = \exp\lbrace -\hat{H}_{0,u}(t) \exp ( \eta_{i,u} ) \rbrace 
\end{equation}    
Because the predictor coefficients are not updated, recalibration will only affect model performance in terms of calibration; discrimination will not change. This method does not accommodate the addition of new predictors to the model.

\subsubsection{Refitting}
Refitting is the most extreme type of updating because the previously estimated predictor coefficients and baseline hazard are discarded. Therefore, it  readily accommodates new predictors. In general, a model could be updated by refitting to new data only or to some combination of old and new data \citep{Hickey2013, Schnellinger2021}. An advantage of refitting to new data only is that the updated model more quickly reflects the new environment but the update may be made on a smaller sample size than the original model development dataset \citep{Davis2019}. Also, if the new data reflects a temporary change in the data generating process or contains data from only a particular segment of the population, the updated model may be overfit to this new data and future performance may be poor \citep{Gama2014}.

\subsubsection{Bayesian dynamic updating}
\label{sec:methodBayes}
Bayesian model updating is a technique for combining knowledge gained in previous models with information in the new data. Applying the Bayesian updating technique for logistic regression from \citet{McCormick2012} to proportional hazards regression and assuming exponentially distributed survival times, the likelihood function can be formulated using a hazard model with an exponential baseline hazard, ie. $h_0(t)= \lambda$. To construct a prior distribution for coefficient $\beta_{j,u}$ of model $M_u$, a normal distribution with parameters derived from the prior period's model $M_{u-1}$ is used. The prior distribution can be written as: $\beta_{j,u} \sim N(\hat{\beta}_{j,u-1}, \hat{\sigma}_{j, u-1}/\xi)$ where $\hat{\beta}_{j,u-1}$ and $\hat{\sigma}_{j,u-1}$ are coefficient $j$'s estimate and uncertainty from the prior period and $\xi < $1 is a `forgetting factor'. The forgetting factor controls the level of uncertainty in the prior; a smaller $\xi$ yields a less informative prior. Survival predictions are then generated from the posterior predictive distribution. Log hazard ratio estimates are taken as the median of the posterior distribution.

\section{Simulation study}
\label{sec:SimStudy}

\subsection{Design}

\subsubsection{Overview and aims}

The goal of this simulation study is to assess methods for dynamic model updating of clinical survival prediction models in changing environments.  Specifically, we aim to identify settings where specific methods outperform others and explore how different types of observational datasets affect the performance of the updating methods. We consider ``open cohort" datasets, where the initial cohort is determined at the start of the study but individuals who have an event are replaced in the cohort with a new member, as well as ``new cohorts" datasets, in which membership will change each period based on receiving a particular treatment or diagnosis.  The simulation scenarios are inspired by our motivating example and thus are characterised by low event rates, predictors with low prevalence in the population, introduction of new treatments and changing baseline risk.

\subsubsection{Data generating mechanisms}

Because the goal is to study the process of updating a previously developed `original' prediction model, we first generated data $D_0$ and fit a model $M_0$ to serve as the starting point. We simulated an initial development dataset with 10,000 individuals for development of the original model. For each individual, 4 covariates were generated: a variable representing age in decades, $X_1 \sim U(1.8, 9.5)$; a prognostic index, $X_2 \sim N(1,1)$; a co-morbidity, $X_3 \sim Bern(p_{X_3})$; and a treatment variable, $X_4 \sim Bern(p_{X_4})$ where $p_{X_3}$ and $p_{X_4}$ vary by scenario. A complete list of parameter values used in the simulation can be found in Web Table 1. Given log hazard ratios $\beta_1, \beta_2, \beta_3, \beta_4$, we generated survival times assuming an exponential distribution with baseline hazard $\lambda$. Survival times were censored at one-year. A Cox proportional hazards model was then fit to this dataset to create the original model. The chosen sample size of 10,000 was determined to be a sufficient size for development of the prediction model \citep{Riley2019}.

We then simulated new data $D_1, \ldots , D_5$, i.e. data arriving after development of the original model, under two different data generating mechanisms: new cohorts and open cohort. In the new cohorts simulation, data is simulated for a new (different) group of individuals each month as might be the case if the data source was all individuals with a positive covid-19 test in that month. In contrast, in the open cohort simulation, follow-up time is added to the existing individuals each month and a small number of new individuals is added to replace those who have had events. This data generation method aims to replicate the use of electronic health records with a cohort start date and then some churn as individuals who are no longer followed up are replaced. Details of the data generation for both styles are found in Web Appendix A. We used $n_{sim}=$600 simulated datasets per scenario.

For both data generating mechanisms, 5 scenarios were studied. Table \ref{tab:scenList} provides a complete list and the associated parameter values for each scenario can be found in Web Table 1. Calibration drift scenarios were characterised by a new baseline hazard in each updated dataset reflecting changes in baseline risk over time. In the ``Rare-1\%" scenario, we assumed 1\% of the total population has a rare risk factor placing them at increased risk of an event and the chance of having this risk factor increased with age. In the context of our motivating example, this could correspond to an age-related comorbidity such as dementia. The treatment indicator, $X_4$, was a time-varying predictor. To simulate the roll-out of a new treatment, such as a vaccine, $p_{X_4}$, the probability of being treated, was increased over time so that an increasingly large segment of the population received treatment. Drawing from the rollout of the covid-19 vaccine in the UK, the population eligible for treatment was based on age with treatment being available to the elderly first and then to successively younger people. In the ``new treatment + comorbidity" scenario, the treatment was made available to those who were immunocompromised ($X_3$=1) or met the minimum age criterion ($X_1$).

\begin{table}
	%	\small
	
	\caption[Listing of all simulation scenarios]{\label{tab:scenRef}Listing of all simulation scenarios and the abbreviated name used in the Results section.}
	\centering
	\small
	\begin{tabular}{ p{.05\linewidth}p{.65\linewidth}p{.16\linewidth}}
		\toprule
		Group & Description & Scenario Name\\
		\midrule
		\multicolumn{3}{l}{Scenarios with calibration drift }  \\
		& Decreasing baseline hazard; event rate decreased from 5\% per year in the first period to 2\% in the last period  &  Decreasing events\\
		& Increasing baseline hazard; event rate increased from 5\% per year in the first period to 8\% in the last period  &  Increasing events\\
		\midrule 
		\multicolumn{3}{l}{Scenario with rare predictor} \\
%		&  Rare risk factor for an event found in 5\% of people & Rare-5\% \\
		& Rare risk factor for an event found in 1\% of people over age 55 & Rare-1\%\\	
		\midrule
		\multicolumn{3}{l}{Scenarios with new predictors} \\
		& Treatment introduced in Q2, made available by age group & New treatment\\
		& Treatment introduced in Q2, made available by age group and to those with a comorbidity (5\% of population)  & {\raggedright New treatment + comorbidity} \\
%		\midrule
%		\multicolumn{3}{l}{Scenarios evaluating impact of forgetting factor} \\
%		& New treatment with forgetting factor = 0.99 & $\xi=$0.99\\
%		& New treatment with forgetting factor = 0.90 & $\xi=$0.90\\
%		& New treatment with forgetting factor = 0.50 & $\xi=$0.50\\
%		& New treatment with forgetting factor = 0.01 & $\xi=$0.01\\
		\bottomrule
		
	\end{tabular}

	\label{tab:scenList}	
\end{table}

\subsubsection{Targets}
The target is the predicted probability of survival to 3 months in the validation data $D_{u+1}$ using a model fit/updated with data from the previous period, $D_u$.

\subsubsection{Methods}

In each simulation scenario, we fit a Cox proportional hazards model (``original model") to an initial dataset with $n$=10,000 observations and 1 year of follow-up time. This model, $M_0$, is the starting point for all updating methods. 

Four strategies, implemented at a single point in time and/or dynamically, were considered for updating the original model over the subsequent 1 year: (1) no update, (2a) refit once at a single time, (2b) refit quarterly, (3a) recalibrate the intercept once at a single time, (3b) recalibrate the intercept quarterly and (4) Bayesian dynamic updating quarterly. In the no update strategy, the original model was retained unchanged and applied to each dataset. Refit refers to fitting a new Cox proportional hazards model to the new data only. With intercept recalibration, the estimated hazard ratios from the original model are retained and only the baseline hazard is re-estimated using the new data. The Bayesian model assumed an exponential baseline hazard. For refit quarterly, recalibrate quarterly and Bayesian update quarterly, the updates are performed dynamically each quarter as new data arrives. 

\begin{figure}
	{\includegraphics[width=6in]{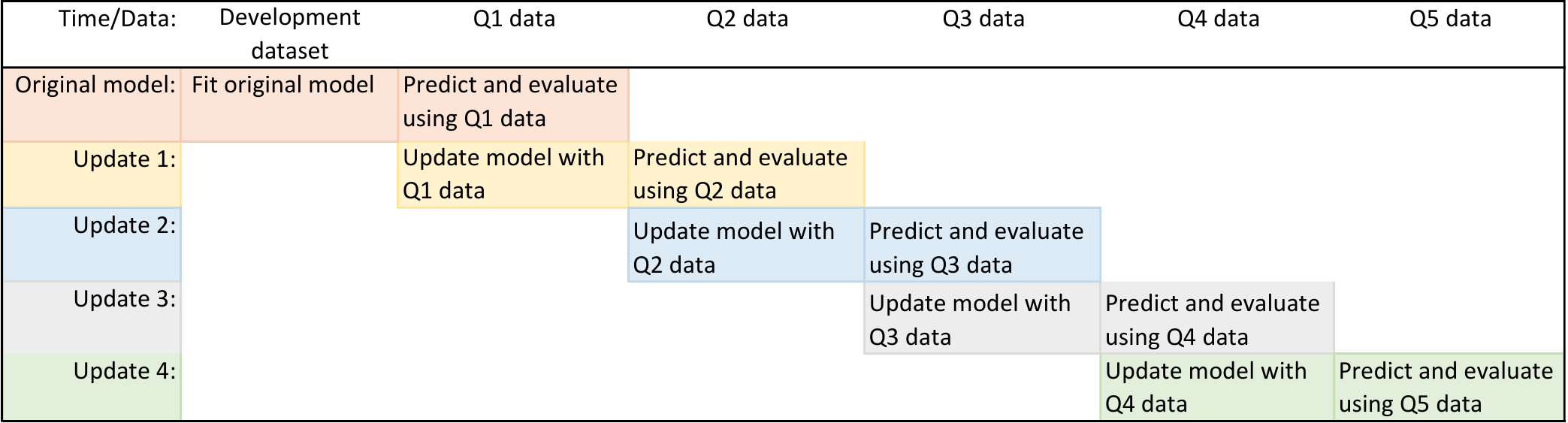}}
	\caption{Illustration of the dynamic updating and evaluation process. Beginning at the top left, an original model was fit to the development dataset and evaluated out-of-sample on the Q1 new data. The model was then updated each quarter with new data and evaluated on the subsequent quarter's data. These updates are called Update 1, 2, 3 and 4 where update $u$ was performed using data from Quarter $u$. A colour version of this figure can be found in the electronic version of the article.}
	\label{fig:UpdateProcess}
\end{figure}

The original model was updated with new data per each updating strategy. Each updated model was then used to predict 3-month survival on the subsequent quarter's data (Q2 - Q5) for evaluation (Figure \ref{fig:UpdateProcess}). In this way, we evaluate predictive performance out-of-sample. We used a quarterly updating cycle for ease of explanation but the updating time frame could be monthly, weekly, daily, or even hourly depending upon the data needs of the updating method. 

In addition to these dynamic strategies with quarterly updating, we also investigate one-time model updating that is not performed according to a pre-determined schedule. Model updating on an ad hoc basis when there are resources available or when the model starts performing poorly is common in practice.  To simulate this, we imagine 7 different analysts faced with the task of updating the original model one time and having to choose when to do that update. We assume 4 of them select the start of a quarter while the remaining 3 select times in between those quarters (drawn randomly); therefore, the analysts will update with 3-months of data beginning at $t=$ 0.0, 0.1, 0.25, 0.46, 0.5, 0.69 and 0.75 months after the start of Year 1. Analysts who begin their update in the middle of a month only have access to the prior 3 months of data as we assume data is collected at the end of the month.

\subsubsection{Performance Measures}
Predictive performance of the updating methods was evaluated using calibration intercept and slope, C-index and Brier score as described in Section \ref{sec:ModAssess}. In the ``Rare-1\%" scenario, we also compare the estimated hazard ratios to the true values used to generate the data.

\subsubsection{Implementation}

All analyses were conducted in R v4.0.2 \citep{RCoreTeam2017}. Survival times were generated using the R package \texttt{simsurv} \citep{Brilleman2021}. We used the \texttt{survival} package \citep{Therneau2000} for Cox proportional hazards regression and the \texttt{pec} package \citep{Mogensen2012} to calculate C-index and Brier score. Calibration intercept and slope were computed using the method described by \citet{Crowson2016}.

Bayesian survival analysis was performed using the \texttt{rstanarm} package \citep{Brilleman2020}. Estimation was via Markov chain Monte Carlo, specifically the No-U-Turn Sampler (Hamiltonian Monte Carlo) implemented in Stan \citep{Hoffman2014}. We used 2 chains, each with 7,500 iterations of which 1,000 were burn-in. This was sufficient to obtain an effective sample size of 10,000 and a Monte Carlo standard error $\approx$  1\% of standard error of the parameter estimates. Convergence was assessed using Gelman and Rubin's Rhat statistic \citep{Gelman1992} with an Rhat $<1.1$ required for convergence. For coefficients not present in the original model (e.g. new treatment) and the rate parameter $\lambda$, we set a prior distribution of N(0, 2.5); all other prior distributions were obtained as described in section \ref{sec:methodBayes}. Typical forgetting factors are just below 1 (e.g. \citet{Raftery2010} chose $\xi$=0.99) and we found that results were not sensitive to the choice of forgetting factor between 0.9 and 0.99 so we chose $\xi$=0.9 (see Web Appendix C for a sensitivity analysis of the forgetting factor).    

If the model could not be refit due to insufficient events/covariate combinations, we retained the model from the previous period to reflect the reality that sometimes it is not possible to refit a model until more data is accrued.

\subsection{Results}
\label{sec:SSResults}

\subsubsection{Scenarios with calibration drift [Decreasing events, Increasing events]}

\begin{figure}
	{\includegraphics[width = 3in]{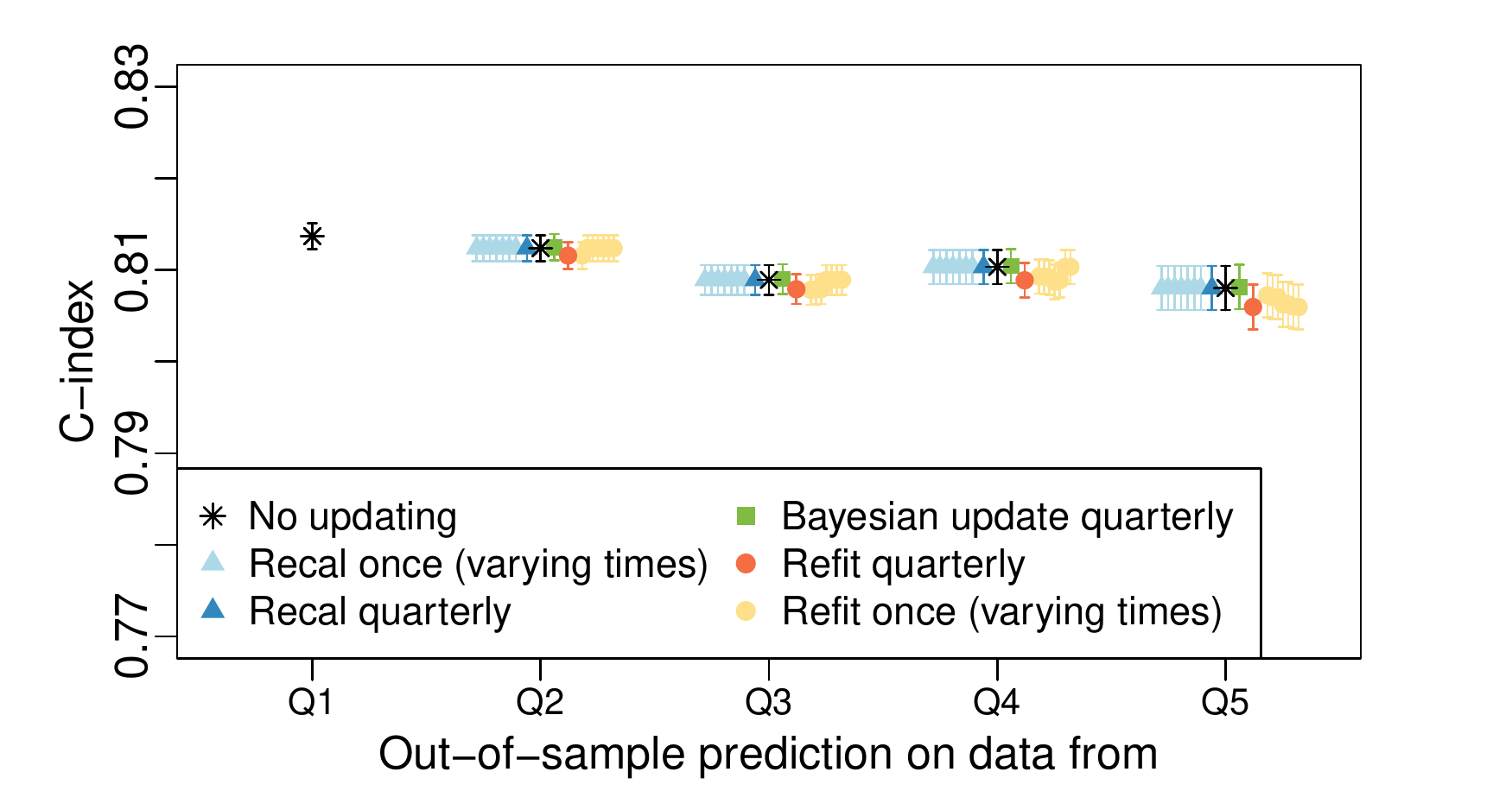}} 
	{\includegraphics[width = 3in]{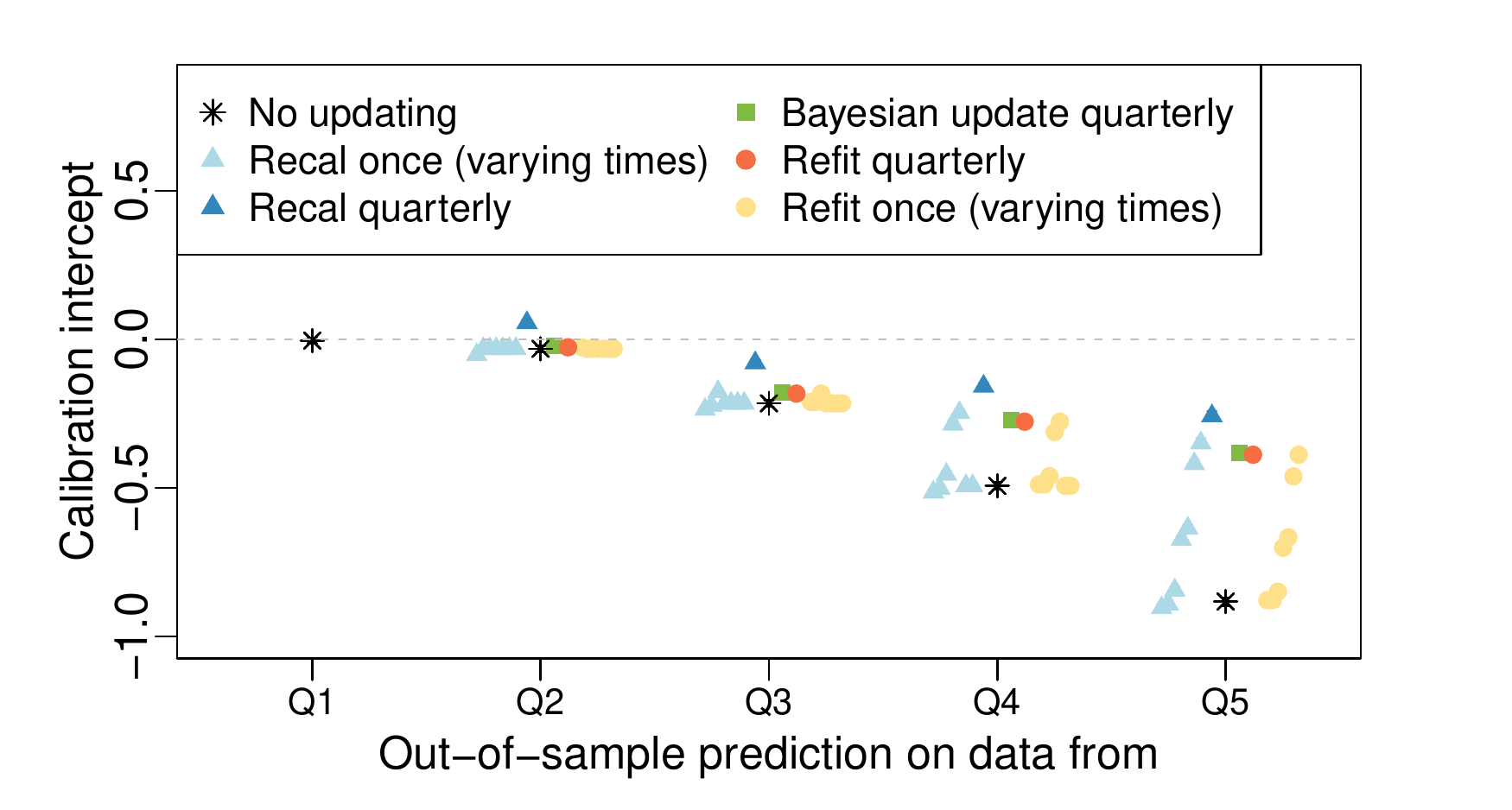}}
	\caption{Open cohort simulation results for a scenario with calibration drift. The left graphic shows the average C-index for each updating method across the 600 simulated datasets at each of the 5 prediction times for a scenario where the event rate decreased over time from 5\% per year in Q1 to 2\% per year in Q5. On the right, the average calibration intercept is shown for the same scenario. Results for `Recal once' and `Refit once' strategies are ordered by update time with the earliest time on the left.   }
	\label{fig:calDrift}
\end{figure}

Discrimination and calibration intercept for the simulation using a decreasing event rate scenario are presented in Figure \ref{fig:calDrift}. Complete results for the calibration drift scenarios can be found in Web Tables 2-3 and Web Figure 2. For the open cohort simulations, in both the increasing and decreasing event rate scenarios, all updating methods (including not updating) produced similar average C-index, Brier score and calibration slope. The average C-index for all methods in the final period was 0.81. Higher values of the C-index indicate better discriminative ability. Target values for calibration intercept and calibration slope are 0 and 1, respectively. Although none of the calibration slopes deviated by more than 0.02 from 1.0, the value of the calibration intercept for a model that was not updated moved further away from zero at each prediction time (see Figure \ref{fig:calDrift}). In the decreasing event rate scenario, updating via quarterly recalibration had the best calibration with average calibration intercepts of  0.06, -0.08, -0.16, -0.26 in Q2-Q5, respectively. Best calibration in the increasing event rate scenario was achieved by the quarterly refitting strategy with average calibration intercepts of 0.10, 0.27, 0.13 and 0.11; Bayesian updating achieved identical calibration intercepts when measured with two significant digits. Models that were recalibrated or refit only once exhibited good calibration at some time points but superior calibration was obtained by a quarterly updating strategy. Results for the new cohorts simulations followed a similar pattern.

\subsubsection{Scenario with rare predictor [Rare-1\%]}

Predictive performance results for the rare predictor scenario are in Web Table 4 and Web Figures 3-4. We considered a scenario, Rare-1\%, where a predictor of survival was relatively uncommon in the population: 1\% of the population had a risk factor for the event and the chance of having the risk factor increased with age. In both simulated dataset styles, open cohort and new cohorts, all updating methods, including no updating, performed similarly for calibration intercept and slope, C-index and Brier score.  

Looking at the coefficient estimates in the open cohort simulation for the rare predictor over time, the average of the original model estimates for the log hazard ratio was 0.76 (MCSE 0.01) compared to the true value of 0.8. The average estimated log hazard ratio for Bayesian updating after each of the 4 updates was 0.78, 0.79, 0.80, 0.81 with MCSE of 0.01 at each time. The refitting strategies showed more variability in the estimates of the log hazard ratio with MCSE of 0.02 at each time and more bias, with log hazard ratio estimates of 0.73, 0.73, 0.75, 0.73 at the 4 update times.  Similarly, in the new cohorts simulation, the refitting strategies exhibited more bias in the estimation of the model coefficients than Bayesian dynamic updating or the original model. Approximately 15\% of the simulated datasets were not able to be refit due to insufficient data.

\subsubsection{Scenarios with new predictors [New treatment, New treatment+comorbidity]}

Full simulation results for the two scenarios with new predictors can be found in Web Tables 5-6 and Web Figures 5-6. In these scenarios, a new treatment was introduced at the beginning of Q2 and rolled out to an increasing percent of the population over time based on age group (``New treatment") or based on age and presence of a comorbidity (``New treatment + comorbidity"). In the new treatment open cohort simulation, Bayesian updating, refitting quarterly and one-time refitting after the introduction of the new treatment offered the best calibration intercept and refitting strategies had the best calibration slope. Recalibration strategies generally had a calibration intercept closer to zero than no updating. All methods had a similar Brier score. Refitting and Bayesian updating strategies showed large improvements in discrimination over no updating or recalibration strategies. The differences in average C-index were 0.03 to 0.07 over no updating in the Q3-Q5 predictions. In the new cohorts simulation, quarterly refitting offered the best calibration intercept but Bayesian updating produced superior calibration slope and discrimination, with a C-index significantly higher than all other methods in Q3-Q5 (Wilcoxon signed rank test p $<$0.001).  

\begin{figure}
	{\includegraphics[width = 3in]{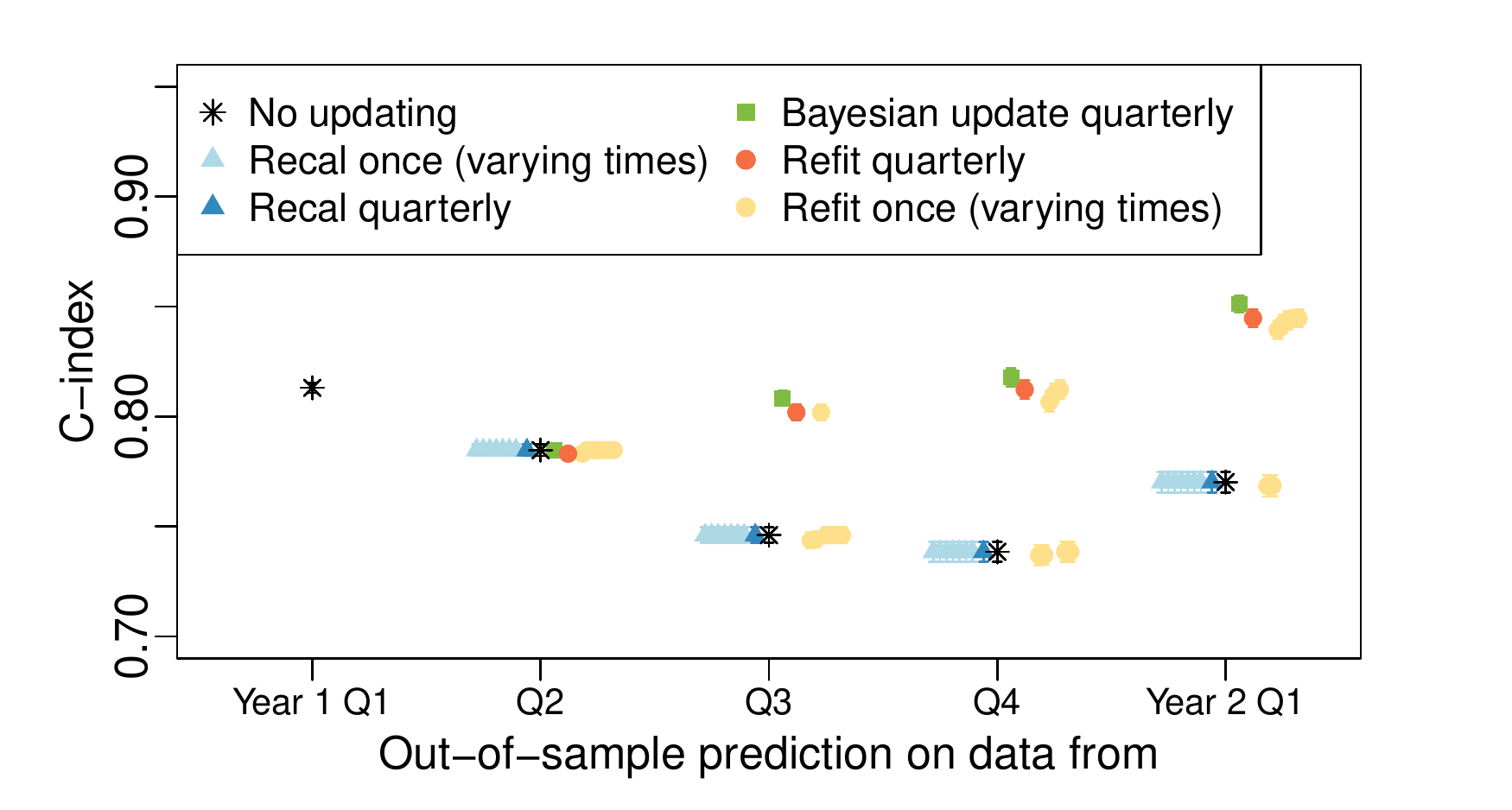}} 
	{\includegraphics[width = 3in]{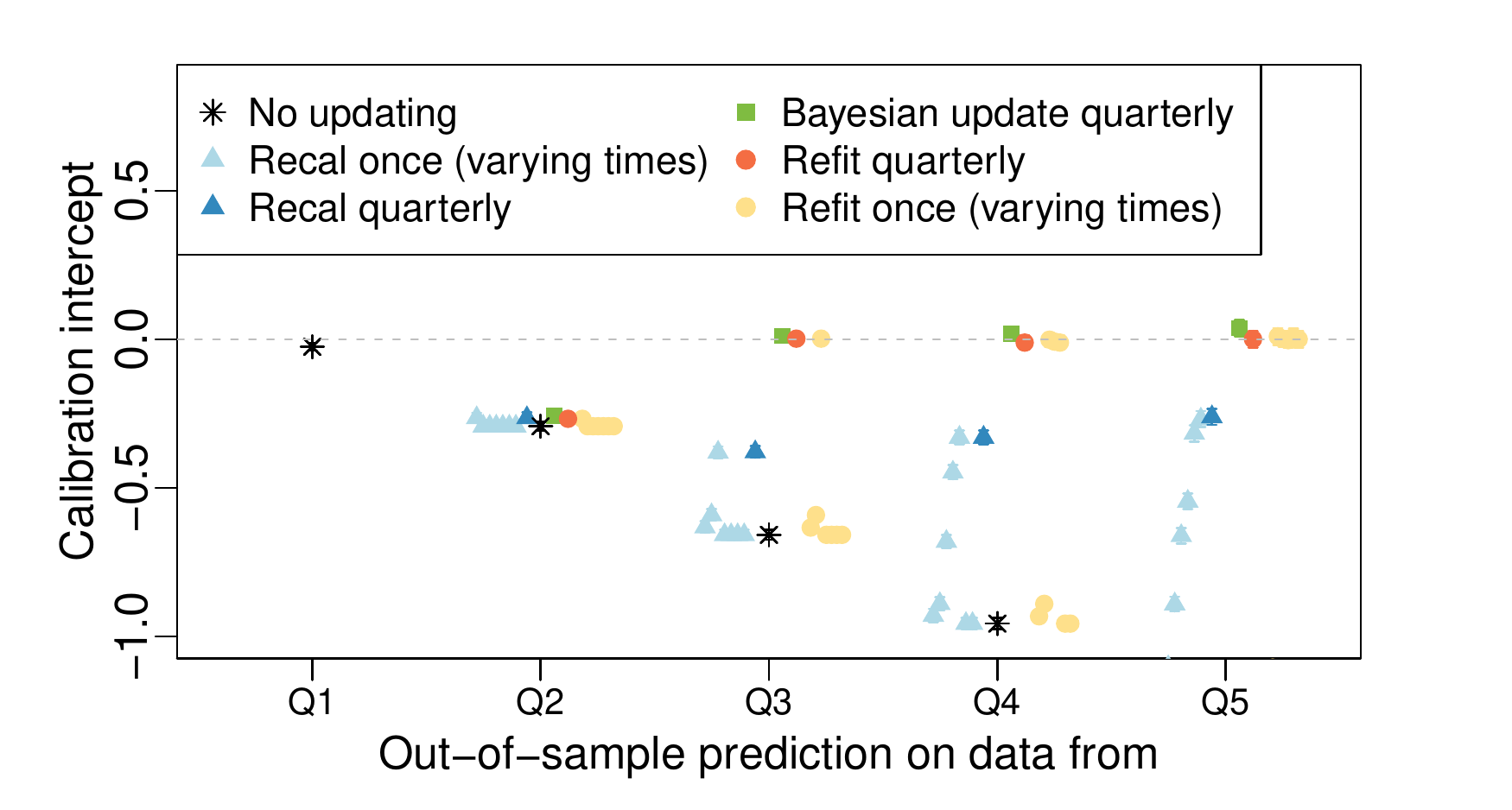}}
	\caption{New cohorts simulation results for the new treatment scenario. The left graphic shows the average C-index for each updating method across the 600 simulated datasets at each of the 5 prediction times for the scenario where a new treatment was introduced at the beginning of Q2. On the right, the average calibration intercept is shown for the same scenario. Results for `Recal once' and `Refit once' strategies are ordered by update time with the earliest time on the left. }
	\label{fig:newTrt2}
\end{figure}

Similar to the new treatment scenario, in the open cohort simulation under the new treatment + comorbidity scenario, Bayesian updating, quarterly refitting and refitting after introduction of the treatment produced the best calibration intercept. Brier scores were similar for all updating methods. Bayesian updating and quarterly refitting produced the best average C-index values (0.82, 0.79, 0.76, 0.76) for the four updates compared to (0.82, 0.79, 0.74, 0.72) for no updating. In the new cohorts simulation, Bayesian updating had the calibration intercept closest to zero for all four updates (-0.13, -0.03, 0.01, -0.04). The highest average C-index came from Bayesian dynamic updating (0.82, 0.82, 0.81, 0.84) with quarterly refitting having the next best discrimination (0.82, 0.80, 0.79, 0.83). In the new treatment + comorbidity scenario, in addition to estimating the effect of the new treatment beginning in Q2, there was also an interaction effect between the new treatment and presence of the comorbidity. As there are small numbers of individuals with the comorbidity who also received the new treatment, in the new cohorts datasets it was not always possible to refit the model at each time point: 8\% (Q2), 35\% (Q3) and 44\% (Q4) models were able to be refit from the 600 simulated datasets.

\section{Updating a covid-19 clinical prediction model}
\label{sec:MotivatingApp}

\subsection{Methods}
\label{sec:exMeths}

To fit and evaluate a covid-19 survival prediction model, data on 1,000,000 individuals aged 18-years or older were obtained from the QResearch database (version 46) for the period 24/01/2020 (the first date where cases were reported in the UK) to 30/04/2021. For consistency with \citet{Hippisley-Cox2021}, our chosen outcome was predicted 70-day survival from covid-19-related death as determined by death certificate information or death within 28 days of a positive covid-19 test. Predictors for each individual were: age, body mass index (BMI), sex, type 1 diabetes, chronic obstructive pulmonary disease (COPD), dementia, and UK region. While this represents a subset of the predictors used in the QCOVID models \citep{Clift2020, Hippisley-Cox2021}, our purpose was not to develop a new covid-19 prediction mode, nor to update an existing one with all their complexities but rather to illustrate methods for model updating. Non-covid-19-related deaths represent a competing risk. Therefore, we fit a sub-distribution hazard model where everyone not experiencing a covid-19 death, (including those who died of other causes) was censored at the end of the study period \citep{Fine1999}. As 18\% of baseline BMI values were missing, we added a missing indicator to the model. Both BMI and age were scaled to have mean of 0 and standard deviation of 1 and then a natural cubic spline was fitted to each with 3 internal knots. These knots were fixed throughout the updating so that previous model estimates would be valid priors for subsequent models. Additionally, 7-day rolling covid-19 case rates by region were included as a predictor \citep{UKCovidDashboard2022} beginning with the first update. We divided the study period into 5 batches of data to create one development dataset followed by four updating datasets as follows: Period 1 (24 Jan - 30 Apr 2020), Period 2 (1 May - 31 Jul 2020), Period 3 (1 Aug - 31 Oct 2020), Period 4 (1 Nov 2020 - 31 Jan 2021) and Period 5 (1 Feb - 30 Apr 2021). An initial sub-distribution hazard model with main effects only was fit to the period 1 dataset with baseline characteristics measured up to the study period start date of 24 January 2020 and follow-up continuing through 30 April 2020. Each subsequent dataset contained information on co-morbidities and regional case rates updated up to the period start date with follow-up continuing up to the end date of the period.  

The three updating methods -- intercept recalibration, refitting and Bayesian updating -- were applied to each updating dataset in succession. The resulting updated prediction models were evaluated out-of-sample on the next period's data for discrimination, calibration and overall predictive performance. For comparison, the original model (without updating) was also evaluated on each updating dataset. Both intercept recalibration and refitting used a proportional hazards model. Bayesian models using a constant baseline hazard were estimated using Stan \citep{Stan2020}. We used 6 chains, each with 3,000 iterations of which 1,000 were burn-in to obtain a Monte Carlo standard error approximately less than or equal to 1\% of standard error of the parameter estimates. Effective sample size was at least 5,000 and Gelman and Rubin's Rhat statistic \citep{Gelman1992} was $<1.01$ indicating convergence. The prior distribution for each coefficient $j$ was $N(\hat{\beta}_{j,t-1}, \hat{\sigma}_{j, t-1}/\xi)$ where $\hat{\beta}_{j,t-1}$ and $\hat{\sigma}_{j,t-1}$ are coefficient $j$'s estimate and uncertainty from the prior period and $\xi$, the forgetting factor, was 0.9.

\subsection{Results}

Web Table 7 presents baseline characteristics of the study population. Table \ref{tab:QResRes} presents performance characteristics for each updating method (recalibrate, refit and Bayesian update) and for the original model without any updating. No single updating method gave best performance across all periods in any of the performance metrics. While refitting produced a higher C-index (0.93) than not updating (0.92) for the first evaluation time, at the second evaluation time refitting had a lower C-index (0.91) than both Bayesian updating (0.92) and no updating (0.94). Refitting was equal to no updating and recalibration (0.91) at the final evaluation time and higher than Bayesian updating (0.90). The C-index for the model after the first Bayesian update (0.76) was the lowest of any method at any time. The Brier scores for all methods in all periods were $<$0.001. Calibration intercepts from Bayesian updating and refitting were closest to 0 for updates using period 2 and 3 data but the original model, without being updated, showed the best calibration intercept in the final updating period.

\begin{table}
		\small
	
\caption{\label{tab:QResRes} Performance of intercept recalibration (Recal), refitting (Refit), and Bayesian dynamic updating (Bayesian) methods to update the prediction model for 70-day covid-19 related death. The original model was fit using data from Period 1 and evaluated using data from period 2. The original model was then updated each period with new data and evaluated using the following period's data. $^\dagger$No update refers to the original model fit in Period 1 and evaluated in each subsequent period without any updating.  }
\centering
\footnotesize

\begin{tabular}{p{.14\linewidth}p{.14\linewidth}p{.05\linewidth}p{.05\linewidth}p{.055\linewidth}p{.05\linewidth}p{.001\linewidth}p{.05\linewidth}p{.05\linewidth}p{.055\linewidth}p{.05\linewidth}}
%\begin{tabular}{{l}{l}{r}{r}{r}{r}{r}{r}{r}{r}{r}}

	\toprule
	
	&&\multicolumn{4}{c}{\textbf{C-index}} && \multicolumn{4}{c}{\textbf{Brier Score}} \\
	
	\midrule

Model fit using	&	Evaluated &	Recal	&	Refit	&	Bayesian	&	No	&	&	Recal	&	Refit	&	Bayesian	&	No	\\
using data from: & using data from: & & & & update$^\dagger$ && & & &update$^\dagger$  \\
\noalign{\vskip 1mm}    
Period 1	&	Period 2	&		&		&		&	0.95	&	&		&		&		&	3E-04	\\
Period 2	&	Period 3	&	0.92	&	0.93	&	0.76	&	0.92	&	&	3E-05	&	3E-05	&	3E-05	&	3E-05	\\
Period 3	&	Period 4	&	0.94	&	0.91	&	0.92	&	0.94	&	&	8E-04	&	7E-04	&	8E-04	&	7E-04	\\
Period 4	&	Period 5	&	0.91	&	0.91	&	0.90	&	0.91	&	&	4E-04	&	4E-04	&	4E-04	&	4E-04	\\

	\noalign{\vskip 1mm}   
	\midrule
		
	&&\multicolumn{4}{c}{\textbf{Calibration intercept}} && \multicolumn{4}{c}{\textbf{Calibration slope}} \\
		
	\midrule

Model fit using	&	Evaluated &	Recal	&	Refit	&	Bayesian	&	No	&	&	Recal	&	Refit	&	Bayesian	&	No	\\
using data from: & using data from: & & & & update$^\dagger$ && & & &update$^\dagger$  \\
\noalign{\vskip 1mm}    
	Period 1	&	Period 2	&		&		&		&	-0.82	&	&		&		&		&	1.10	\\
	Period 2	&	Period 3	&	-0.79	&	0.30	&	0.15	&	-2.12	&	&	0.92	&	0.82	&	0.86	&	0.92	\\
	Period 3	&	Period 4	&	3.16	&	-0.04	&	-0.01	&	0.56	&	&	0.93	&	0.90	&	0.86	&	0.93	\\
	Period 4	&	Period 5	&	-0.66	&	-1.10	&	-1.12	&	-0.49	&	&	0.85	&	0.89	&	0.88	&	0.85	\\

\bottomrule	
	
\end{tabular}

\end{table}

\section{Discussion}

In this study, we investigated the performance of discrete and dynamic updating methods for clinical survival prediction models. Overall, the dynamic updating strategies at regular intervals outperformed a single update. We found that Bayesian dynamic updating offered the best performance in the simulation study in situations with new predictors and less data and that all methods generally improved calibration. In our motivating example, where the environment was changing rapidly and the outcome was rare, no single updating method outperformed the others.

Although no one method performed best in all circumstances, we can draw several conclusions from the study. First, intercept recalibration is an effective tool for calibration maintenance that requires little data, is not computationally intensive and will not change the rank order of predicted survival probabilities between two individuals. This simplifies reporting of the updated model and may be less confusing for users.  Second, although refitting can produce a good performing model when adequate data is available, it does not outperform Bayesian updating in general, even when there are new predictors.  And, because refitting requires a large number of observations/events and it may produce abruptly changing hazard ratio estimates over time, refit models have the greatest chance of being overfit and are the most likely to produce substantially different predictions for an individual compared to the previous model. \citet{Riley2021} cautions that attempts to ameliorate this overfitting by applying shrinkage techniques may be unreliable due to estimation uncertainty of the tuning parameters and is best used with a larger sample size.

Bayesian dynamic updating offers the advantages of both recalibration and refitting and, in the simulation study, was the best performer in the majority of scenarios and time points across the evaluation criteria. However, it is the most computationally intensive updating method we studied with a single update taking 12-24 hours using the QResearch dataset of 1,000,000 records. Although the detail of how a Bayesian model is estimated would be complicated to explain to non-statisticians, we believe the concept of a Bayesian update -- that it combines knowledge from the current model with information from new data -- is intuitively appealing. The relatively stable hazard ratio estimates are a further advantage and may help engender trust amongst the users about the updating process. 

In the covid-19 application, the discrimination of the first Bayesian updated model was poor. This was primarily due to the method for obtaining the priors for the first update. The original model was fit using a Cox proportional hazards model and the first priors were taken from this model. This mimics the situation where the analyst(s) updating the model are different from the analyst(s) who originally developed it and they may not have access to the original development dataset. In this case, the fitted model is the only source of information for the priors. The poor performance in the first update occurred because the Bayesian updating assumed exponentially distributed survival times whereas the original model made no such assumption and, therefore, the coefficients were estimated with a different baseline hazard. Had the actual survival times been exponentially distributed (as in the simulation study) obtaining the priors from a Cox fit would have produced the same priors as those from an exponential model. We recommend care in obtaining priors when access to the development data is not possible. Also, more complex models of the baseline hazard could be used in the Bayesian model but these come with increased computational cost.

The results from the illustration of updating a model that predicts catching and dying from covid-19 were inconclusive. Different updating strategies, including no updating, performed well at different times with different metrics. During the study period, the UK experienced multiple waves of covid-19. Therefore, a model updated during a time of high prevalence could be tested out-of-sample at a time of low prevalence and good calibration was difficult to maintain. We hypothesise that a model predicting risk of covid-19 related death in those with a positive covid-19 test would be less susceptible to these cycles. Also, our example was constructed using a 3-month updating period to ensure sufficient events in each dataset to allow for refitting. However, a monthly updating cycle using Bayesian dynamic updating may have performed better. Interesting future work would be to update a covid-19 prediction model including the period that saw the introduction of vaccines in the UK.

In the simulation study, refitting performed better in the open cohort-style data scenarios than in those with new cohorts. This result is due to the smaller amount of data available in the new cohorts dataset and the fact that, in many cases, we were unable to fit a new model to the new data only. Analysts may consider using some combination of old and new data to overcome the lack of new data but how much old versus new data to include is subjective and the choice can have a strong impact on the refit prediction model. Also, we must be aware of the patient mix in the new dataset and whether it differs from that of the original development data and the data we expect to see in the future. Bayesian dynamic updating is well-suited for this case as the model can be successively updated with new evidence without waiting for a dataset as large as the original development dataset to accumulate.

We powered the simulation study to detect a difference of 0.01 in the C-index but it is difficult to know how big a difference in each of the performance criteria would be clinically significant. For example, we found evidence of statistically different values of the C-index and Brier score based on a Wilcoxon signed rank test even when the values themselves were identical to two significant digits. It is unlikely that these differences are relevant in a practical sense. Also, when event rates are very low, as in the covid-19 application, Brier scores may not be informative for model comparison because all methods are likely to predict a low event probability for most individuals. When averaged, these small differences overwhelm the few cases where predicted risk is higher and the resulting Brier scores will be small.

The choice of updating method should be carefully selected considering the available data, existence of new predictors and subject matter knowledge. Equally important, however, is the development of a strategy for ongoing dynamic updating including the recurring collection of new data to capture environment changes and distributional shifts. A plan for communicating the updated model and refreshing web pages and calculators is also required when implementing a dynamic strategy. We also wish to caution against over-automating the updating process as clinical input may identify trends and environmental changes before they are evident in the data.

\section{Acknowledgements}

This study is funded by the National Institute for Health Research (NIHR) following a commission by Department of Health and Social Care. RHK is supported by a UKRI Future Leaders Fellowship (MR/S017968/1). KDO is funded by a Royal Society-Wellcome Trust Sir Henry Dale fellowship, grant number  218554/Z/19/Z.

We acknowledge the contribution of EMIS practices who contribute to QResearch® and EMIS Health and the the Universities of Nottingham and Oxford for expertise in establishing, developing and supporting the QResearch database. QRESEARCH® is a registered trademark of Egton Medical Information Systems Limited and the University of Nottingham. We used an extract of 1 million patients from the QResearch database (version 46). We also used the linked ONS death registry. The Office of National Statistics bears no responsibility for the analysis or interpretation of the data.

%\clearpage

\bibliography{DynUpdate}

\section*{Supporting Information}
Web Appendices, Tables, and Figures referenced in Sections \ref{sec:SimStudy} and \ref{sec:MotivatingApp} are available with this paper. Sample code for the updating methods and a Stan model file are available at \url{https://github.com/KamTan/DynamicModelUpdate}. Although we are not permitted to make the QResearch data public, interested researchers may apply for access to the QResearch database using an online application system. Details can be found at: \url{https://www.qresearch.org/information/information-for-researchers/}. 

\end{document}